\documentclass[a4paper,11pt]{article}
\usepackage{pos}

\usepackage{physics}
\usepackage{xcolor}
\usepackage{subcaption}
\newcommand{\bea}{\begin{eqnarray}}
\newcommand{\eea}{\end{eqnarray}}
\usepackage{mathtools}
\usepackage{cancel} 
\newcommand{\monec}{\raisebox{-0.43\totalheight}{\includegraphics[scale=0.50]{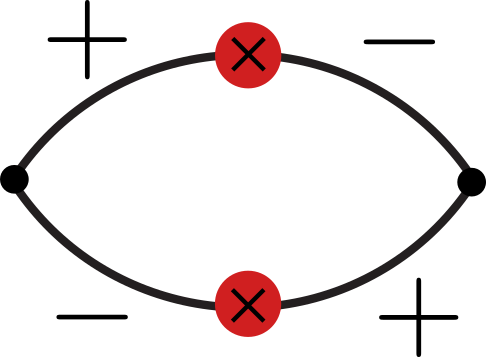}}}
\newcommand{\moned}{\raisebox{-0.43\totalheight}{\includegraphics[scale=0.50]{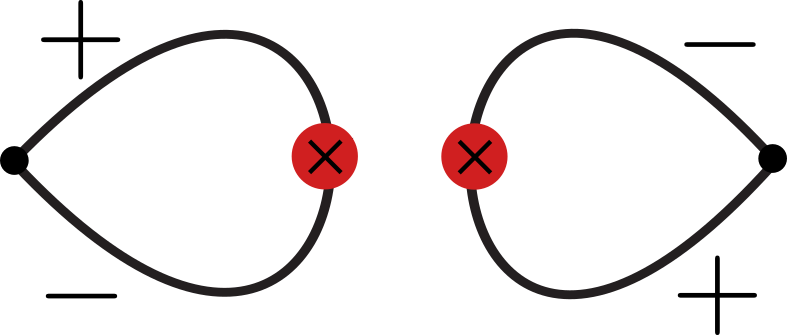}}}
\newcommand{\mtwoc}{\raisebox{-0.43\totalheight}{\includegraphics[scale=0.50]{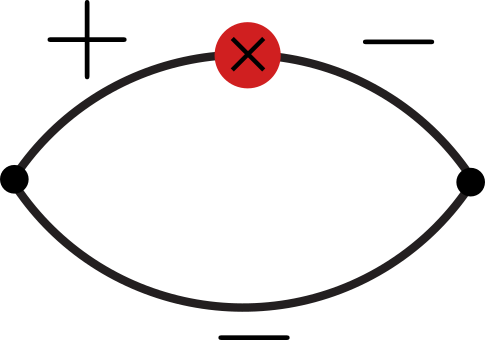}}}
\newcommand{\mtwod}{\raisebox{-0.43\totalheight}{\includegraphics[scale=0.50]{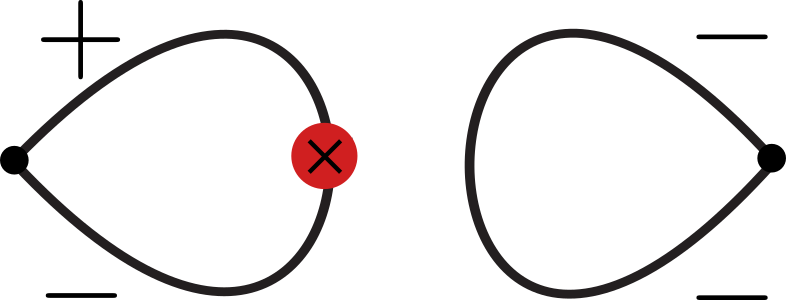}}}
\newcommand{\Mconnpm}{\raisebox{-0.4\totalheight}{\includegraphics[scale=.40]{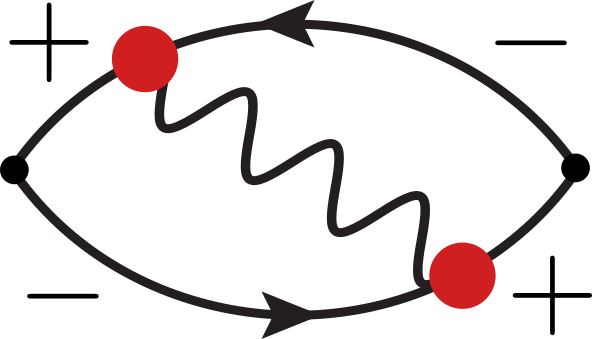}}}
\newcommand{\Mdiscpm}{\raisebox{-0.2\totalheight}{\includegraphics[scale=.40]{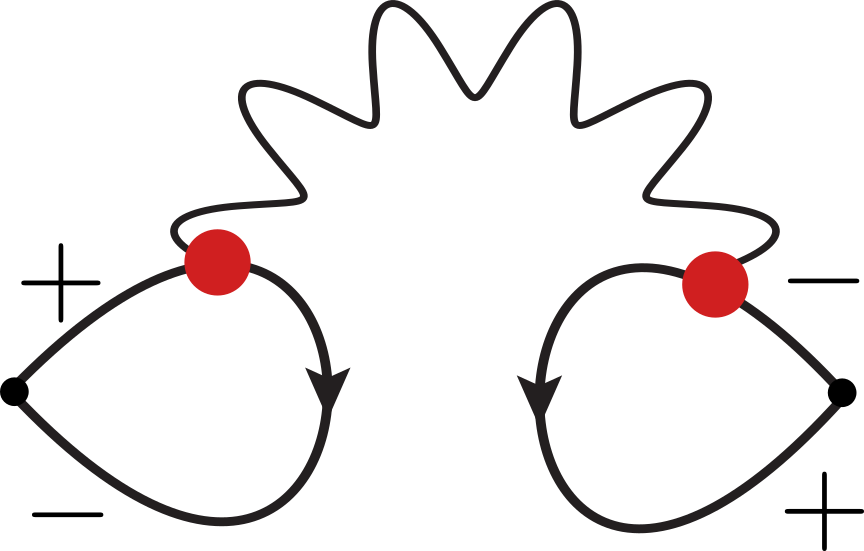}}}

\title{Lattice determination of the pion mass difference $M_{\pi^{+}} - M_{\pi^{0}}$ at order $\mathcal{O}(\alpha_{em})$ and $\mathcal{O}( (m_{d}-m_{u})^{2})$ including disconnected diagrams.}
\ShortTitle{Lattice determination of $M_{\pi^{+}}- M_{\pi^{0}}$ at $\mathcal{O}(\alpha_{em})$ and $\mathcal{O}( (m_{d}-m_{u})^{2})$.}
\author[a]{R. Frezzotti}
\author*[b]{G. Gagliardi}
\author[b,c]{V. Lubicz}
\author[d]{G. Martinelli}
\author[b]{F. Sanfilippo}
\author[b]{S. Simula}

\affiliation[a]{Dipartimento di Fisica and INFN, Università di Roma “Tor Vergata”,\\
Via della Ricerca Scientifica 1, I-00133 Rome, Italy}

\affiliation[b]{Istituto Nazionale di Fisica Nucleare, Sezione di Roma Tre, \\
Via della Vasca Navale 84, I-00146 Rome, Italy}

\affiliation[c]{Dipartimento di Fisica, Università Roma Tre, \\
Via della Vasca Navale 84, I-00146 Rome, Italy}

\affiliation[d]{ Dipartimento di Fisica and INFN Sezione di Roma La Sapienza, \\ Piazzale Aldo Moro 5, I-00185 Rome, Italy}

\emailAdd{roberto.frezzotti@roma2.infn.it}
\emailAdd{giuseppe.gagliardi@roma3.infn.it}
\emailAdd{vittorio.lubicz@uniroma3.it}

\emailAdd{guido.martinelli@roma1.infn.it}

\emailAdd{francesco.sanfilippo@infn.it}

\emailAdd{silvano.simula@roma3.infn.it}

\abstract{
We present our preliminary results concerning the charged/neutral pion mass difference $M_{\pi^{+}} - M_{\pi^{0}}$ at order $\mathcal{O}(\alpha_{em})$ in the QED interactions, and for $M_{\pi^{+}} - M_{\pi^{0}}$  at order $\mathcal{O}\left( (m_{d}-m_{u})^{2}\right)$ in the strong isospin-breaking term. The latter contribution provides a determination of the $\rm{SU}(2)$ chiral perturbation theory low-energy constant $\ell_{7}$, whose present estimate is affected by a rather large uncertainty. The disconnected contributions appearing in the diagrammatic expansion of $M_{\pi^{+}} - M_{\pi^{0}}$, being very noisy, are notoriously difficult to evaluate and have been neglected in our previous calculations. By making use of twisted mass Lattice QCD simulations and adopting the RM123 method, we will show that taking profit from our recently proposed \textit{rotated twisted-mass} (RTM) scheme, tailored to improve the signal on these kinds of observables, it is possible to evaluate the disconnected diagrams with good precision. For the QED induced pion mass difference, we obtain, after performing the extrapolation towards the continuum and thermodynamic limit and at the physical point, the preliminary value $M_{\pi^{+}}-M_{\pi^{0}} = 4.622~(95)~{\rm MeV}$, that is in good agreement with the experimental result. For the determination of the low-energy constant $\ell_{7}$, our result $\ell_{7} = 2.5~(1.4)\times 10^{-3}$, which is limited so far to a single lattice spacing, is in agreement and improves phenomenological estimates.

}

\FullConference{%
 The 38th International Symposium on Lattice Field Theory, LATTICE2021
  26th-30th July, 2021
  Zoom/Gather@Massachusetts Institute of Technology
}

\begin{document}
\maketitle

\section{Introduction}
In the last decade, the precision achieved in the computation of several observables relevant for flavour physics by lattice QCD has reached a level where electromagnetic and strong isospin-breaking (IB) effects can no longer be neglected~\cite{Aoki:2021kgd}. Among the many phenomenologically relevant observables for which the calculation of IB effects are important, in this presentation we are concerned with the calculation of the charged/neutral pion mass difference $M_{\pi^+}-M_{\pi^0}$, at leading order in the electromagnetic interactions $\left(\mathcal{O}\left(\alpha_{em}\right)\right)$, and at order $\mathcal{O}((m_{d}- m_{u})^{2})$ in the QCD IB. The latter, which is subdominant with respect to the leading QED contribution of $\mathcal{O}(\alpha_{em})$, turns out to be important for the evaluation of the $\rm{SU}(2)$ low-energy costant (LEC) $\ell_{7}$, which parameterizes strong IB effects in the chiral perturbation theory (ChPT) Lagrangian at next-to-leading order (NLO), and whose uncertainty is currently larger than $50\%$~\cite{di_Cortona_2016, Frezzotti:2021ahg}. The computation of $\ell_{7}$ can be performed also using an alternative method, in the following denoted as the "matrix element method", in which $\ell_{7}$ is determined from the coupling of the neutral pion $\pi^{0}$ to the isoscalar operator $P^{0}= \left(\bar{u}\gamma_{5}u + \bar{d}\gamma_{5} d\right)/\sqrt{2}$, and which is presented here as well and compared with the "meson mass method". \\

Our strategy is based on the RM123 approach~\cite{de_Divitiis_2012,deDivitiis:2013xla} in which the lattice path-integral is expanded in powers of the small parameters $\alpha_{em}$ and $m_{d}-m_{u}$, with $\alpha_{em} \sim (m_{d}-m_{u})/\Lambda_{QCD}\sim \mathcal{O}(10^{-2})$. This approach allows to express the expectation value of any given observable in QCD$+$QED as a power series in $\alpha_{em}$ and  $m_{d}-m_{u}$ whose coefficients are related to correlation functions evaluated in the isospin symmetric theory. We perform the diagrammatic expansion of the relevant correlation functions using a recently proposed scheme for twisted-mass (TM) regularization of lattice QCD, the \textit{rotated twisted-mass} (RTM) scheme~\cite{Frezzotti:2021slr}, which has been shown to be convenient for lattice calculations of IB effects, especially within the RM123 approach. In particular, the quark disconnected diagrams which appear in the diagrammatic expansion of the pion mass splitting at both $\mathcal{O}(\alpha_{em})$ and $\mathcal{O}( (m_{d}-m_{u})^{2})$, as well as the disconnected diagram contributing to the coupling of the neutral pion $\pi^{0}$ to the isoscalar operator  $P^{0}$ at order $\mathcal{O}(m_{d}-m_{u})$, which are notoriously very noisy in standard TM, are affected by much smaller statistical errors if evaluated in the RTM scheme~\cite{Frezzotti:2021slr}.\\

The calculation of the charged/neutral pion mass difference at $\mathcal{O}(\alpha_{em})$ has been performed using the pure QCD isosymmetric gauge ensembles generated by the Extended Twisted Mass Collaboration (ETMC) with $N_{f}=2+1+1$ dynamical quarks~\cite{Baron:2010bv,baron2011light, Carrasco:2014cwa}. After extrapolating to the physical pion mass and to the continuum and infinite volume limit, our preliminary result is:
\begin{align}
\label{eq:prelim_res_QED}
M_{\pi^{+}}-M_{\pi^{0}} = 4.622~(64)_{stat.}(70)_{syst.}~{\rm MeV},\qquad  \left[\rm{PDG}: 4.5936(5)~{\rm MeV}\right]~.
\end{align}
Concerning instead the determination of $M_{\pi^{+}}-M_{\pi^{0}}$ at  $\mathcal{O}( (m_{d}-m_{u})^{2})$, and the determination of $\langle 0 | P^{0} | \pi^{0}\rangle$ at $\mathcal{O}( m_{d}-m_{u})$, we made use of the $N_{f}=2+1+1$ gauge configurations produced with Wilson-clover TM fermions by the ETMC~\cite{alexandrou2021ratio, alexandrou2021quark}. For this pilot study, we limited our simulations to a single value of the lattice spacing $a\simeq 0.095~{\rm fm}$ and to a single pion mass $M_{\pi}\simeq  260~{\rm MeV}$. Our final estimate of the value of $\ell_{7}$, which has been also presented in Ref.~\cite{Frezzotti:2021ahg}, is
\begin{align}
\label{eq:res_l7}
\ell_{7} = 2.5(1.3)_{stat.}(0.5)_{syst.}\times 10^{-3} = 2.5(1.4)\times 10^{-3}~,
\end{align}

\section{Evaluation of the LEC $\ell_{7}$}
\label{sec:II}
According to the analysis of Refs.~\cite{Gasser:1983yg,GASSER1985465,Frezzotti:2021ahg}, the value of the LEC $\ell_{7}$ can be determined using two different methods, that we denote as the \emph{mass} and the \emph{matrix element} methods. The first relies on the fact that $\ell_{7}$ parametrizes the charged/neutral pion mass difference induced by QCD IB through 
\begin{align}
\label{l7_M1_definition}
    \ell_7 =~2\frac{\left(M_{\pi^+}-M_{\pi^0}\right)_{QCD}}{(m_{u}-m_{d})^{2}}\cdot\frac{m_{\ell}^2 f_{\pi}^2}{M_\pi^3}\,,\qquad (\textrm{mass method})
\end{align}
where $(M_{{\pi}^{+}} - M_{\pi^{0}})_{QCD}$ indicates only the pure QCD contribution to the pion mass difference, $m_{\ell}$ is the light quark mass, $f_{\pi}$ the pion decay constant, and $M_{\pi}$ is the pion mass in isosymmetric QCD. \\

In the matrix element method, instead, one exploits the fact that for $m_{u} \ne m_{d}$,  the neutral pion has a non vanishing iso-singlet component, which is quantified by the matrix element
\begin{align}
Z_{P^{0}\pi^{0}} &\equiv \langle 0 | P^{0} | \pi^{0}\rangle =  \frac{1}{\sqrt{2}}\langle 0 | \left( \bar{u}\gamma^{5}u + \bar{d}\gamma^{5}d\right) | \pi^{0} \rangle~.
\end{align}
The value of $Z_{P^{0}\pi^{0}}$ at $\mathcal{O}(m_{d}-m_{u})$ is then proportional to $\ell_{7}$, which in turn can be determined through the relation~\cite{Frezzotti:2021ahg} 
\begin{align}
\label{l7_M2_definition}
\ell_{7} = -\frac{Z_{P^{0}\pi^{0}}}{m_{u}-m_{d}}\cdot \frac{f_{\pi}m_{\ell}^{2}}{M_{\pi}^{4}}\,,\qquad (\textrm{matrix element method})~,
\end{align} \\

Within the RM123 approach, both $Z_{P^{0}\pi^{0}}$ and $(M_{{\pi}^{+}} - M_{\pi^{0}})_{QCD}$ are computed treating the IB term in the QCD action, which is proportional to the mass difference $\Delta m=(m_d-m_u)/2$, as a small perturbation. We perform the RM123 expansion in the RTM scheme, in which the Lagrangian of the light doublet $\psi'_{\ell} = (u',d')$, is given by~\cite{Frezzotti:2021slr}
\begin{align}
\label{RTM}
\mathcal{L}_{RTM}(\psi'_{\ell}) = \bar{\psi}'_{\ell}(x)\left[ \gamma_{\mu}\widetilde{\nabla}_{\mu}  -i\gamma_{5}\tau_{3}W(m_{cr})+ m_{\ell}  \right]\psi'_{\ell}(x) + \Delta m\mathcal{L}_{IB}(x)~,
\end{align}
where $\widetilde \nabla_\mu$ is the lattice symmetric covariant derivative, while the critical Wilson term $W(m_{cr})$, which includes the critical mass, and the IB term $\mathcal{L}_{IB}$, are given by
\begin{equation}
W(m_{cr}) = - a\, \frac{r}{2}\, \nabla_\mu \nabla^*_\mu + m_{cr}(r)\,,\qquad \mathcal{L}_{IB}(x)=\bar{\psi}'_{\ell}(x) \tau_{1}\psi'_{\ell}(x)~.
\end{equation}
In the RTM scheme, the IB term $\mathcal{L}_{IB}$, being proportional to $\tau^{1}$, is flavour-changing w.r.t. the quark fields $u',d'$ which are regularized in Eq.~(\ref{RTM}) with opposite values of the Wilson parameter $r=\pm 1$, and are related to the physical up and down quark fields $u,d$ through the rotation
\begin{align}
\begin{pmatrix}
u \\
d 
\end{pmatrix}
= \frac{1}{\sqrt{2}}\begin{pmatrix}
1 & 1 \\
-1 & 1
\end{pmatrix}
\begin{pmatrix}
u' \\
d' 
\end{pmatrix}~,
\end{align}
The mass difference $(M_{\pi^{+}} - M_{\pi^{0}})_{QCD}$ and the matrix element $\langle 0 | P^{0} |  \pi^{0} \rangle$, can be extracted respectively from the physical correlators $C_{\pi^{+}\pi^{+}}(t) - C_{\pi^{0}\pi^{0}}(t)$ and $C_{P^{0}\pi^{0}}(t)$ ($C_{AB}(t) = \langle 0 | A(t) B^{\dag}(0) | 0 \rangle)$, that in term of the \textit{rotated fields} of the RTM basis are given by
\begin{align}
\label{C1}
C_{\pi^{+}\pi^{+}}(t) - C_{\pi^{0}\pi^{0}}(t) = -2\,C_{\pi'^{+}\pi'^{-}}(t)~,\quad
C_{P^{0}\pi^{0}}(t) =  -\frac{1}{\sqrt{2}}\left[ C_{P'^{0}\pi'^{+}}(t) + C_{P'^{0}\pi'^{-}}(t)\right] ~,
\end{align}
where $
\pi'^{-}=\bar{u}'\gamma_{5}d', ~\pi'^{+}=\bar{d}'\gamma_{5}u',~ P'^{0}=\left[\bar{u}'\gamma_{5} u' + \bar{d}'\gamma_{5} d'\right]/\sqrt{2}$.\\

In turn, the RM123 expansion of the correlators appearing in the r.h.s. of Eq.~(\ref{C1}), respectively at second and first order in $\Delta m$, and obtained using the RTM Lagrangian Eq.~(\ref{RTM}), is given by (see~\cite{Frezzotti:2021slr} for a detailed derivation)
\begin{align}
\label{M1_expansion}
&\begin{aligned} 
C_{\pi^{+}\pi^{+}}(t) - C_{\pi^{0}\pi^{0}}(t)  &= -2\,C_{\pi'^{+}\pi'^{-}}(t)\\[1pt]
&= -2\left(\frac{Z_{S}}{Z_{P}}\right)^{2}(\Delta m)^{2}~\bigg[\,\,\underbrace{\monec}_{C_{MM}^{conn.}(t)}\,\, - \,\, \underbrace{\moned}_{C_{MM}^{disc.}(t)} \,\,\bigg]
\end{aligned} \\[-4pt]
\label{M2_expansion}
&\begin{aligned} 
C_{P^{0}\pi^{0}}(t)  &= -\frac{1}{\sqrt{2}}\left( C_{P'^{0}\pi'^{+}}(t) + C_{P'^{0}\pi'^{-}}(t)\right) \\[1pt]
&= -2\frac{Z_{S}}{Z_{P}}\Delta m ~\bigg[\,\, \underbrace{\mtwoc}_{C_{MEM}^{conn.}(t)}\,\, - \,\, \underbrace{\mtwod}_{C_{MEM}^{disc.}(t)}\,\, \bigg]
\end{aligned}
\end{align}
where the black lines  represent the isosymmetric light quark propagators with Wilson parameter $r=\pm 1$, as indicated on each quark line, black vertices denote the insertion of $\gamma_{5}$, and red vertices denote the insertion of the identity matrix corresponding to the perturbation $\mathcal{L}_{IB}$. Finally, in Eqs.~(\ref{M1_expansion}) and~(\ref{M2_expansion}) we included the renormalization constant (RC) of the operator $\mathcal{L}_{IB}$ and of the mass difference $m_{d}-m_{u} = 2\Delta m$ which, in our twisted mass formulation, are given respectively by $Z_{S}$ and $Z_{P}^{-1}$. \\

From the correlator $C_{\pi^{+}\pi^{+}}(t) - C_{\pi^{0}\pi^{0}}(t)$, the pion mass difference $(M_{\pi^{+}}-M_{\pi^{0}})_{QCD}$ at $\mathcal{O}(\Delta m^{2})$ can be computed using (see e.g. Ref.~\cite{Frezzotti:2021ahg})
\begin{align}
\frac{C_{\pi^{+}\pi^{+}}(t) - C_{\pi^{0}\pi^{0}}(t)}{C_{\pi\pi}^{\rm{\rm{isoQCD}}}(t)} = \textrm{const.} + (M_{\pi^{+}}- M_{\pi^{0}})_{QCD}\cdot(T/2-t)\cdot\tanh{\left[ M_{\pi}(T/2-t)\right]} + \ldots~,    
\end{align}
where the dots represent subleading exponentials, $T$ is the lattice time extent, and $M_{\pi}$ is the ground state mass extracted from the isosymmetric pion correlator $C_{\pi\pi}^{\rm{isoQCD}}$, represented by the single connected diagram without any mass insertion, and computed with opposite values of the Wilson parameter $r$. Similarly, the matrix element $\langle 0 | P^{0} | \pi^{0} \rangle$ at $\mathcal{O}(\Delta m)$, can be evaluated using
\begin{align}
\frac{C_{P^{0}\pi^{0}}(t)}{C_{\pi\pi}^{\rm{isoQCD}}(t)} = \frac{Z_{P^{0}\pi^{0}}}{Z_{\pi\pi}} +\ldots ~, \end{align}
where $Z_{\pi\pi} \equiv \langle \pi | P^{\dag}_{\pi} | 0 \rangle $ is the overlap between the isoQCD pion state and the interpolating source $P_{\pi}$, and the dots represent again subleading exponentials.
\subsection{Numerical results for $\ell_{7}$}
\begin{figure}
\begin{subfigure}[t]{.5\textwidth}
  \includegraphics[width=0.98\linewidth]{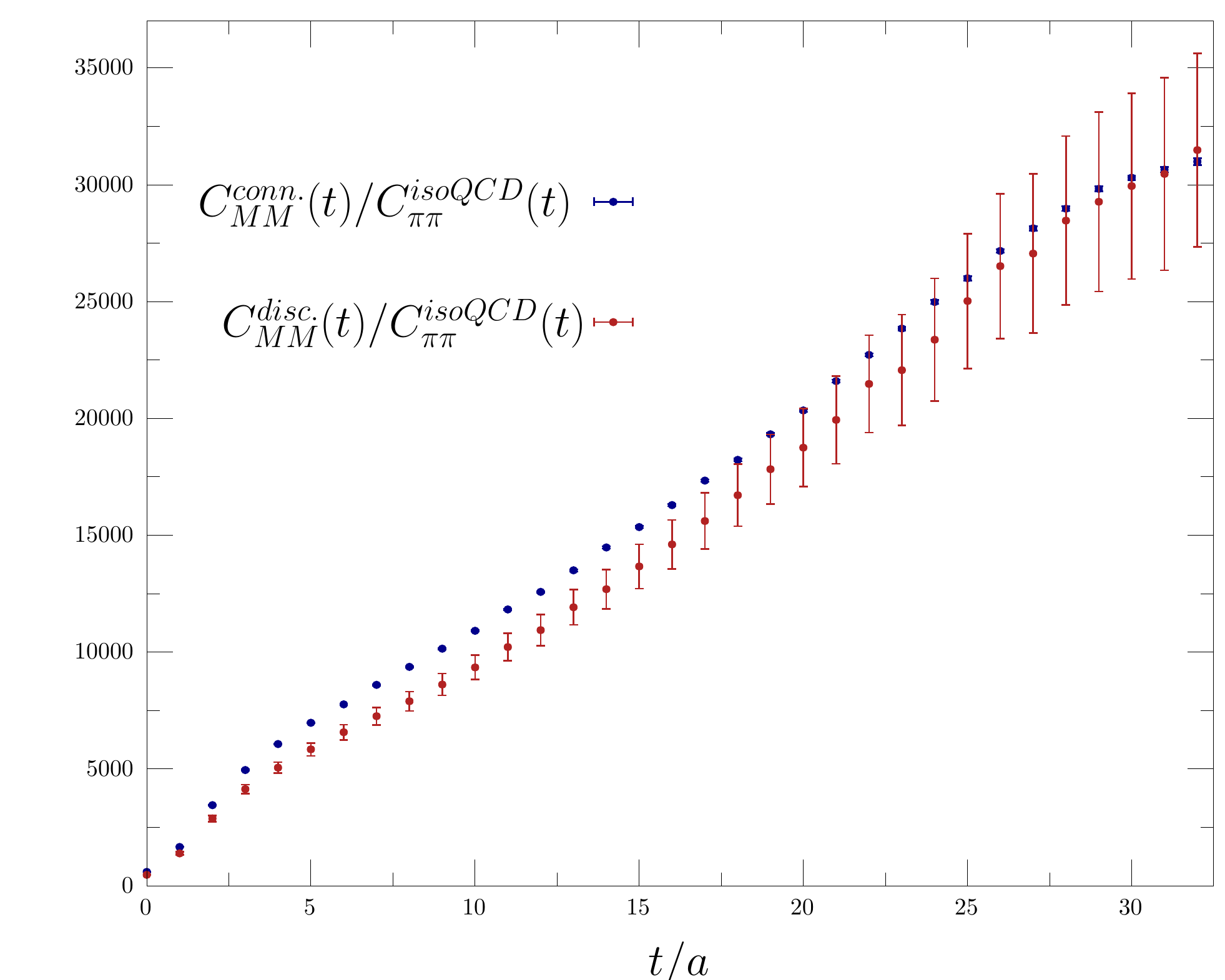}  
 
\end{subfigure}
\begin{subfigure}[t]{.5\textwidth}
  \includegraphics[width=0.98\linewidth]{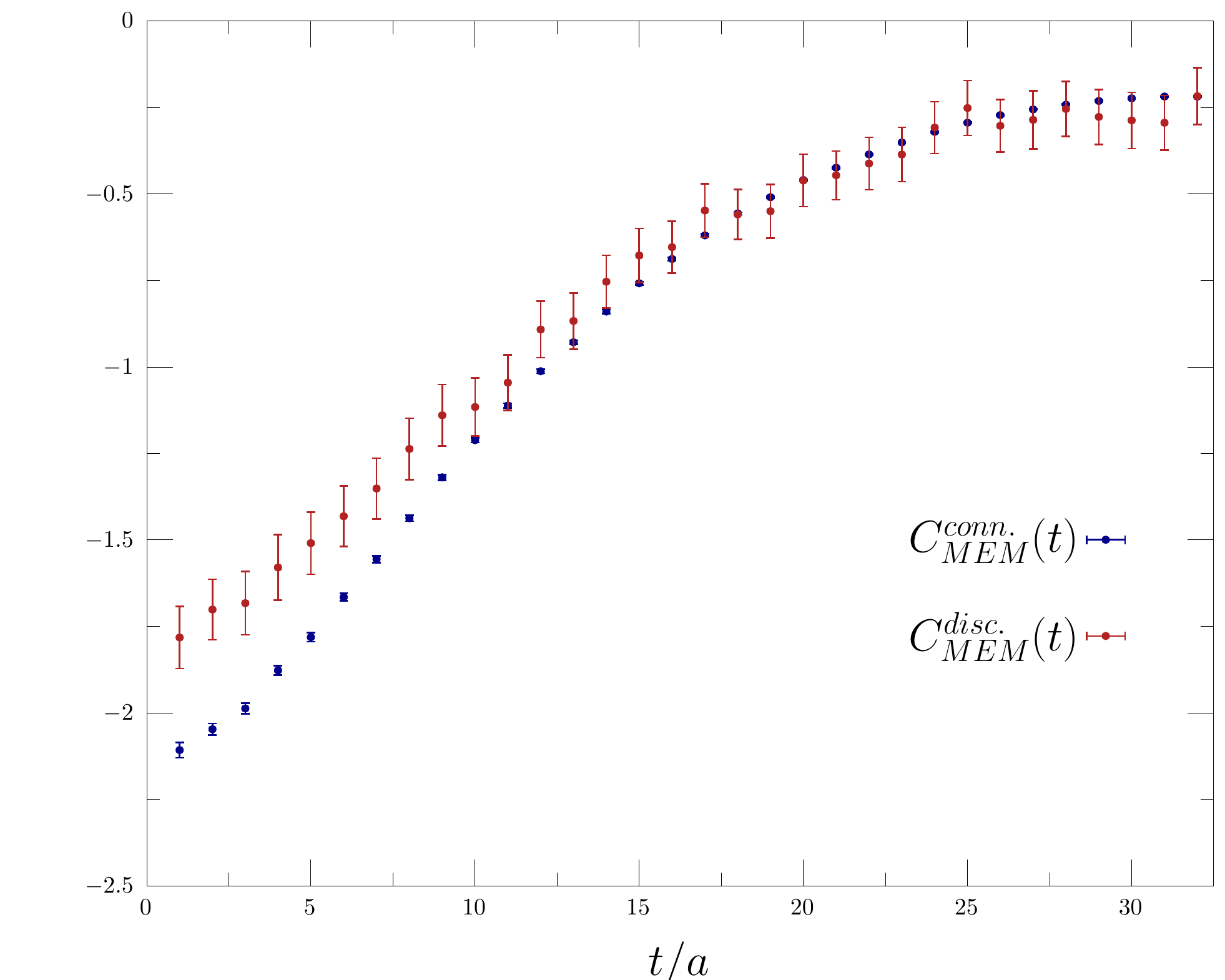}  

\end{subfigure}
\caption{\small\it Comparison between the connected and disconnected diagrams appearing in the mass method (Left plot) and in the matrix element method (Right plot).}
\label{fig1}
\end{figure}
\begin{table}
    \centering
    \begin{tabular}{c| c c c c c}
    \hline
        Ensemble  & $\hat{m}_{\ell}$ & $\hat{f}_{\pi}$ & $\hat{M}_{\pi}$ &  $Z_{P}/Z_{S}$   \\
        \hline
    cA211.30.32     &  0.0030 & 0.06674 (15)  & 0.12530 (16) & 0.726 (3) \\
    \hline
    \end{tabular}
    \caption{\small\it List of the input parameters used for the determination of $\ell_{7}$ on the cA211.30.32 ensemble.}
    \label{tab1}
\end{table}

\begin{figure}
    \centering
    \includegraphics[width= 0.56\linewidth]{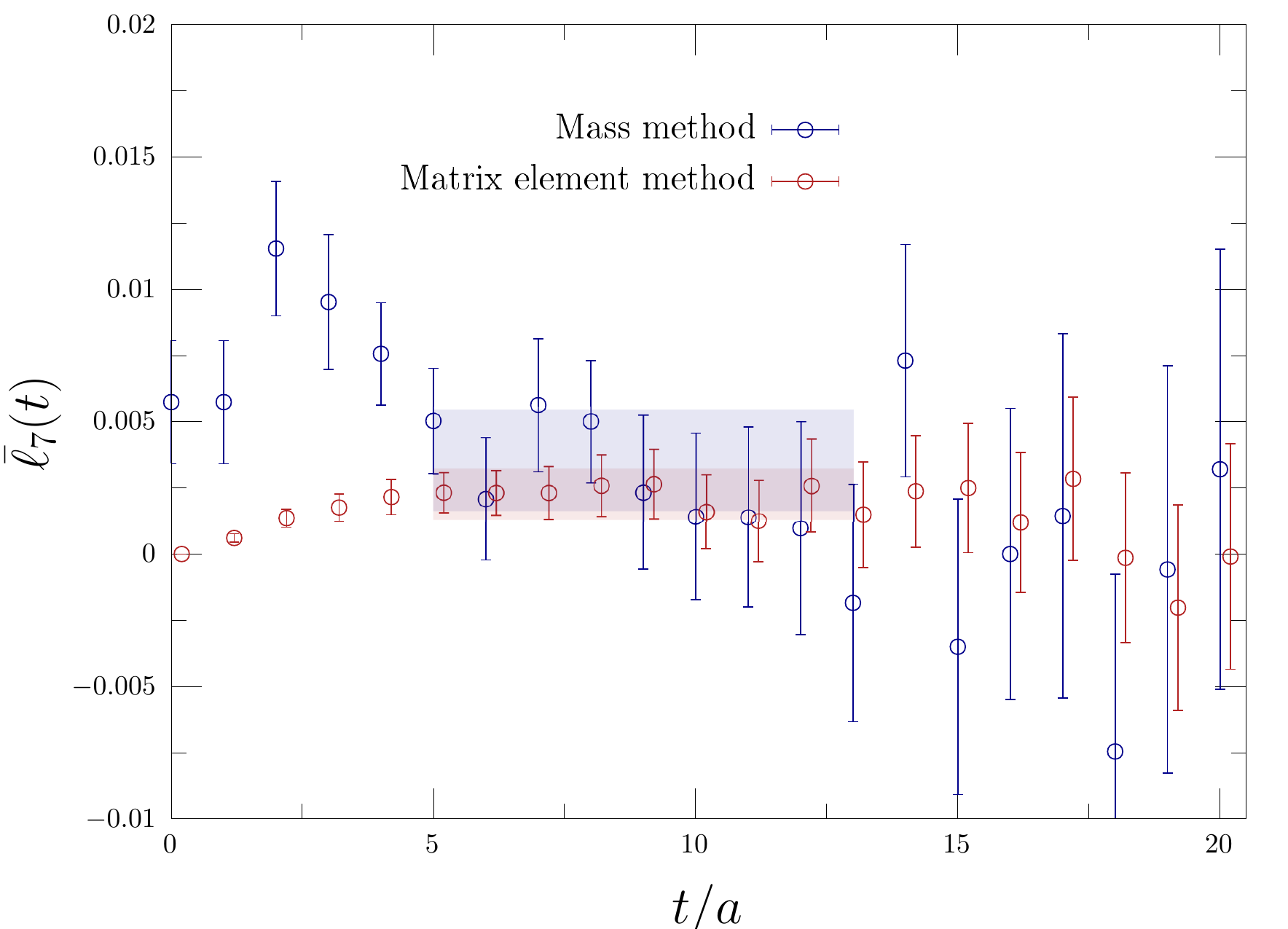}
    \caption{\small\it Determination of $\ell_{7}$ on the cA211.30.32 ensemble using both the mass method and the matrix element method. The two bands correspond to the result of a constant fit in the time interval $[5,13]$.}
    \label{fig.3}
\end{figure}

For this exploratory study of $\ell_{7}$ we limited the simulations to a single ensemble (cA211.30.32), generated by the ETMC with Wilson-clover TM fermions. In Tab.~\ref{tab1} we collected the values of the input parameters that have been used for the determination of $\ell_{7}$. Relying on Eqs.~(\ref{l7_M1_definition}) and~(\ref{l7_M2_definition}), we have built the following estimators to extract $\ell_{7}$ from the diagrams in Eqs.~(\ref{M1_expansion}) and~(\ref{M2_expansion}):
\begin{align}
\label{estimator_M1}
\bar{\ell}_{7}(t) &= \left(\frac{Z_{S}}{Z_{P}}\right)^{2}\cdot\frac{\hat{f}_{\pi}^{2}\hat{m}_{\ell}^{2}}{\hat{M}_{\pi}^{3}}\cdot \partial_{t}\left[\frac{ C_{MM}^{conn.}(t) - C_{MM}^{disc.}(t)}{C_{\pi\pi}^{\rm{isoQCD}}(t)}\right]\quad\,\,\,\,\,\,\,\,\,\,\,\,\,\, \left(\text{mass method}\right)~,\\[10pt]
\label{estimator_M2}
\bar{\ell}_{7}(t) &= -\left(\frac{Z_{S}}{Z_{P}}\right)\cdot\frac{ \hat{f}_{\pi}\hat{m}_{\ell}^{2}}{\hat{M}_{\pi}^{4}}\cdot \hat{Z}_{\pi\pi}\cdot\left[\frac{ C_{MEM}^{conn.}(t) - C_{MEM}^{disc.}(t)  }{ C_{\pi\pi}^{\rm{isoQCD}}(t)} \right]\quad\left(\text{matrix element method}\right)~,
\end{align}
where $\hat{Z}_{\pi\pi} = \hat{f_{\pi}}\hat{M}_{\pi}\sinh{\hat{M}_{\pi}}/2\hat{m}_{\ell}$, and the operator $-\partial t$ corresponds to the evaluation of the so-called effective slope defined as
\begin{align}
\label{eq:def_meff}
\delta m_{eff}(t) \equiv -\partial_{t} \frac{\delta C(t)}{C(t)} &= \frac{\left(\frac{\delta C(t)}{C(t)} -\frac{\delta C(t-1)}{C(t-1)}\right)}{(\frac{T}{2}-t)\tanh{(M(\frac{T}{2}-t))} - (\frac{T}{2}-t+1)\tanh{(M(\frac{T}{2}-t+1))}} ~ ,
\end{align}   
where in the mass method $C(t)= C_{\pi\pi}^{\rm{isoQCD}},~\delta C(t) = C_{MM}^{conn.}(t) - C_{MM}^{disc.}(t)$. The estimators tend to $\ell_{7}$ in the large time limit $t/a \gg 1$. In the panel of Fig.~\ref{fig1} we show our determination of the diagrams $C_{MM}^{conn.}(t)$ and $C_{MM}^{disc.}(t)$ normalized over the isosymmetric pion correlator, along with our determination of $C_{MEM}^{conn.}(t)$ and $C_{MEM}^{disc.}(t)$. As it can be seen, the signal of $\ell_{7}$ comes in both cases from a delicate cancellation between connected and disconnected contributions, which makes this calculation a non-trivial task given that a very good precision on both diagrams is needed in order to obtain a good signal-to-noise ratio in the difference. In this respect the use of the RTM is essential to improve the signal. In Fig.~\ref{fig.3}, we show instead the estimators of Eqs.~(\ref{estimator_M1}) and~(\ref{estimator_M2}). The matrix element method shows smaller statistical errors w.r.t. the mass method, and both estimators are consistent in the plateaux region. In both cases the signal disappears into noise at $t\sim 15$, and $\ell_{7}$ can be extracted through a costant fit at smaller times only. We fitted both estimators in the time interval $[5,13]$ and obtained in this way 
\begin{align}
\label{eq:l7_final_M1_M2}
\ell_{7} = 3.5~(2.0)\times 10^{-3}~ \left(\text{MM}\right)~,\qquad
\ell_{7} = 2.3~(1.0)\times 10^{-3}~ \left(\text{MEM}\right)~.   
\end{align}
Even if our analysis is limited to a single value of the lattice spacing, the difference between the two determinations in Eq.~(\ref{eq:l7_final_M1_M2}) can be used as a first (likely conservative) estimate of the systematic error associated to the missing continuum extrapolation, given that they are affected by different $\mathcal{O}(a^{2})$ lattice artifacts. Moreover, the systematic associated to the missing chiral extrapolation $m_{\ell}\to 0$ is expected to be small as compared to our statistical error, given that in ChPT the presence of a non-zero $m_{\ell}$ corresponds to a NNLO correction to Eqs.~(\ref{l7_M1_definition}) and~(\ref{l7_M2_definition}). Making use of Eqs.~(38)-(43) of Ref.~\cite{alexandrou2021quark} to combine the two determinations, we get
\begin{align}
\label{eq:final_l7}
\ell_{7} = 2.5~(1.3)_{stat.}(0.5)_{syst.}\times 10^{-3} = 2.5~(1.4)\times 10^{-3}~,
\end{align}
which is in agreement but significantly improves the phenomenological estimate of Ref.~\cite{di_Cortona_2016} ($7~(4) \times 10^{-3}$), and the determination of the RBC/UKQCD Collaboration~\cite{Boyle_2016} ($6.5~(3.8)\times 10^{-3}$). 
\section{Pion mass difference at $\mathcal{O}(\alpha_{em})$}
The (UV finite) difference between charged and neutral pion correlators at order $\mathcal{O}(\alpha_{em})$, is obtained in the RTM scheme~(see~\cite{Frezzotti:2021slr} for more details on this point) from the double insertion in the isospin symmetric pion correlator of the iso-triplet component of the electromagnetic current

\begin{align}
\label{eq:Jib}
J_{\mu}^{ib}(x)=  \Delta q\bar{\psi}'_{\ell}(x)\tau_{1}\gamma_{\mu}\psi'_{\ell}(x)~,
\end{align}
where $\Delta q = (q_u -q_d)/2$, which, in the RTM basis, induces a mixing between the $u'$ and $d'$ quarks. Performing the corresponding Wick contractions, one gets  that the difference between charged and neutral pion correlators at $\mathcal{O}(\alpha_{em})$ can be expressed in the RTM scheme as (see again~\cite{Frezzotti:2021slr} for a derivation)
\begin{align}
\label{corr_RTM_QED}  
C_{\pi^{+} \pi^{+}}(t) - C_{\pi^{0}\pi^{0}}(t)= -2C_{\pi'^{+}\pi'^{-}}(t) = -2e^{2}(\Delta q)^{2}Z_{A}^{2}\,\bigg[\,\,\underbrace{\Mconnpm}_{\delta\bar{C}_{\pi}^{exch.}(t)}\,\,-\,\, \underbrace{\Mdiscpm}_{\delta\bar{C}_{\pi}^{disc.}(t)}\,\,\bigg]~,
\end{align}
where the red vertices represent the insertion of $J_{\mu}^{ib}$, and we showed again explicitly the sign of the Wilson parameter on each quark line, which gets always flipped at the e.m. vertex where the $u^{\prime}$ quark turns into a $d^{\prime}$ quark and viceversa. In the previous expression $Z_{A}$ is the RC of the axial current which in our TM setup renormalizes the local current $J_{\mu}^{ib}$. Instead, at order $\mathcal{O}(m_{d}- m_{u})$,  $C_{\pi'^{+}\pi'^{-}}(t)=0$ since a single insertion of $\mathcal{L}_{IB}$ cannot convert a $\pi'^{+}$ into a $\pi'^{-}$. This is expected since pion correlators are symmetric with respect to the exchange between up and down quarks. From Eq.~(\ref{corr_RTM_QED}), it follows that the pion mass difference at $\mathcal{O}(\alpha_{em})$ is given in the RTM scheme by
\begin{align}
\label{eq:pion_mass_QED}
M_{\pi^{+}} - M_{\pi^{0}} = 2e^{2}(\Delta q)^{2} Z_{A}^{2} \partial_{t} \frac{\delta \bar{C}_{\pi}^{exch.}(t) - \delta \bar{C}_{\pi}^{disc.}(t)}{C_{\pi\pi}^{\rm{isoQCD}}(t)}~,   
\end{align}
where the operator $-\partial_{t}$ is defined as in Eq.~(\ref{eq:def_meff}) with $\delta C(t) = \delta \bar{C}_{\pi}^{exch.}(t)-\delta \bar{C}^{disc.}_{\pi}(t)$. To cope with the infrared divergence of the photon propagator (wiggly lines in Eq.~(\ref{corr_RTM_QED})), we adopt the $QED_{L}$ regularization and set $A_{\mu}(k_{0}, \vec{k}=0)=0$ for all $k_{0}$~\cite{deDivitiis:2013xla}.

\subsection{Numerical results for $M_{\pi^{+}}- M_{\pi^{0}}$ at $\mathcal{O}(\alpha_{em})$}
For this study, we made use of the first set of $N_{f}=2+1+1$ ensembles of Wilson TM fermions generated by the ETMC. The ensembles correspond to pion masses in the range $M_{\pi} \in [200~{\rm MeV}, 500~{\rm MeV}]$ and lattice spacings from $a\sim 0.088~{\rm fm}$ down to $a\sim 0.062~{\rm fm}$. Detailed informations on the ensembles are provided in Ref.~\cite{Carrasco:2014cwa}. For the RC $Z_{A}$ appearing in Eqs.~(\ref{corr_RTM_QED}) and~(\ref{eq:pion_mass_QED}), we made use of the more precise determination obtained from the method $M_{2}$ of Ref.~\cite{Carrasco:2014cwa}. Moreover, we only considered the subset of the ETMC ensembles corresponding to $M_{\pi}L \geq 3.8$ to limit the presence of QCD exponential finite size effects (FSEs). To improve the precision on the disconnected diagram of Eq.~(\ref{corr_RTM_QED}), we devised a new numerical technique, tailored for quark disconnected diagrams, in which the photon propagator is evaluated exactly by working in momentum space, and therefore the statistical noise coming from its stochastic representation is absent. The method, which will be illustrated in details in a forthcoming publication, combined with the benefit of the RTM scheme, allowed us to obtain an uncertainty of order $\mathcal{O}(1\%)$ on the value of the disconnected diagram. \\

We found it useful to consider the dimensionless ratio
\begin{align}
   R_{\pi} \equiv \frac{M^{2}_{\pi^+} -M^{2}_{\pi^0}}{f_{\pi}^{2}} \approx  \frac{2M_{\pi}}{f_{\pi}^{2}}\left( M_{\pi^{+}} - M_{\pi^{0}}\right)~,
\end{align}
after applying to both $f_{\pi}$ and $M_{\pi}$, the $\rm{SU}(2)$ ChPT finite volume corrections at NNLO $+$ resummation, i.e. the Colangelo-D{\"u}rr-Haefeli (CDH) formulae~\cite{Colangelo:2005gd}. The latter depend on the knowledge of the four, scale dependent, $\rm{SU}(2)$ LECs $\bar{\ell}_{1}, \bar{\ell}_{2}, \bar{\ell}_{3}, \bar{\ell}_{4}$, and in this work we adopt the same choice made in Ref.~\cite{alexandrou2021ratio} for the values of the LECs.\\
\begin{figure}
    \centering
    \includegraphics[scale=0.50]{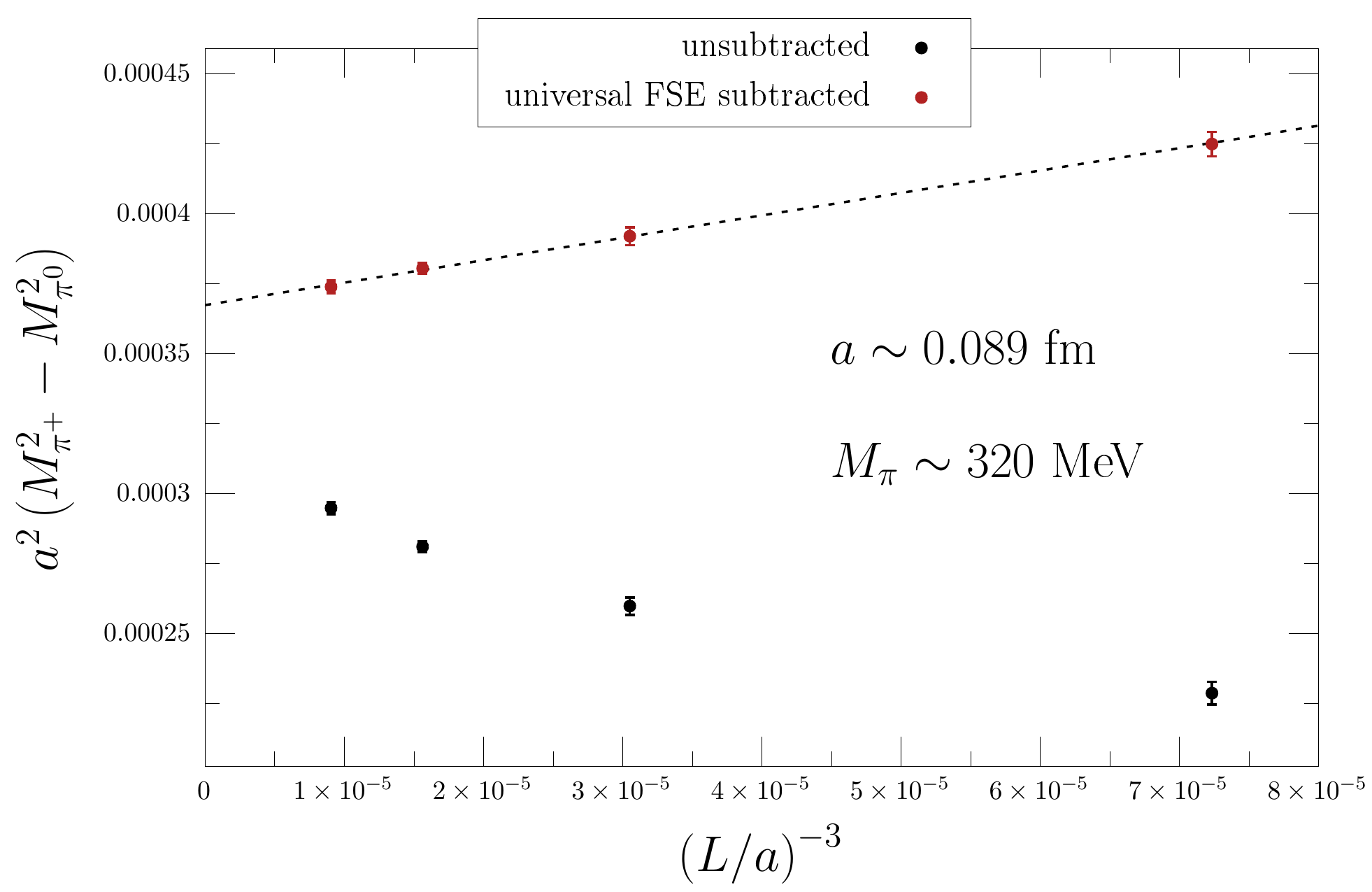}
    \caption{\small\it The (squared) pion mass difference in the RTM scheme, for the ensembles of type A40.XX, which share a common value of the pion mass ($M_{\pi}\simeq 320~\textrm{MeV}$) and of the lattice spacing, but differ in the lattice extent $L$. The dashed line is the result of a linear fit in $(L/a)^{-3}$.}
    \label{fig:QED1}
\end{figure}

\begin{figure}
    \centering
    \includegraphics[scale=0.50]{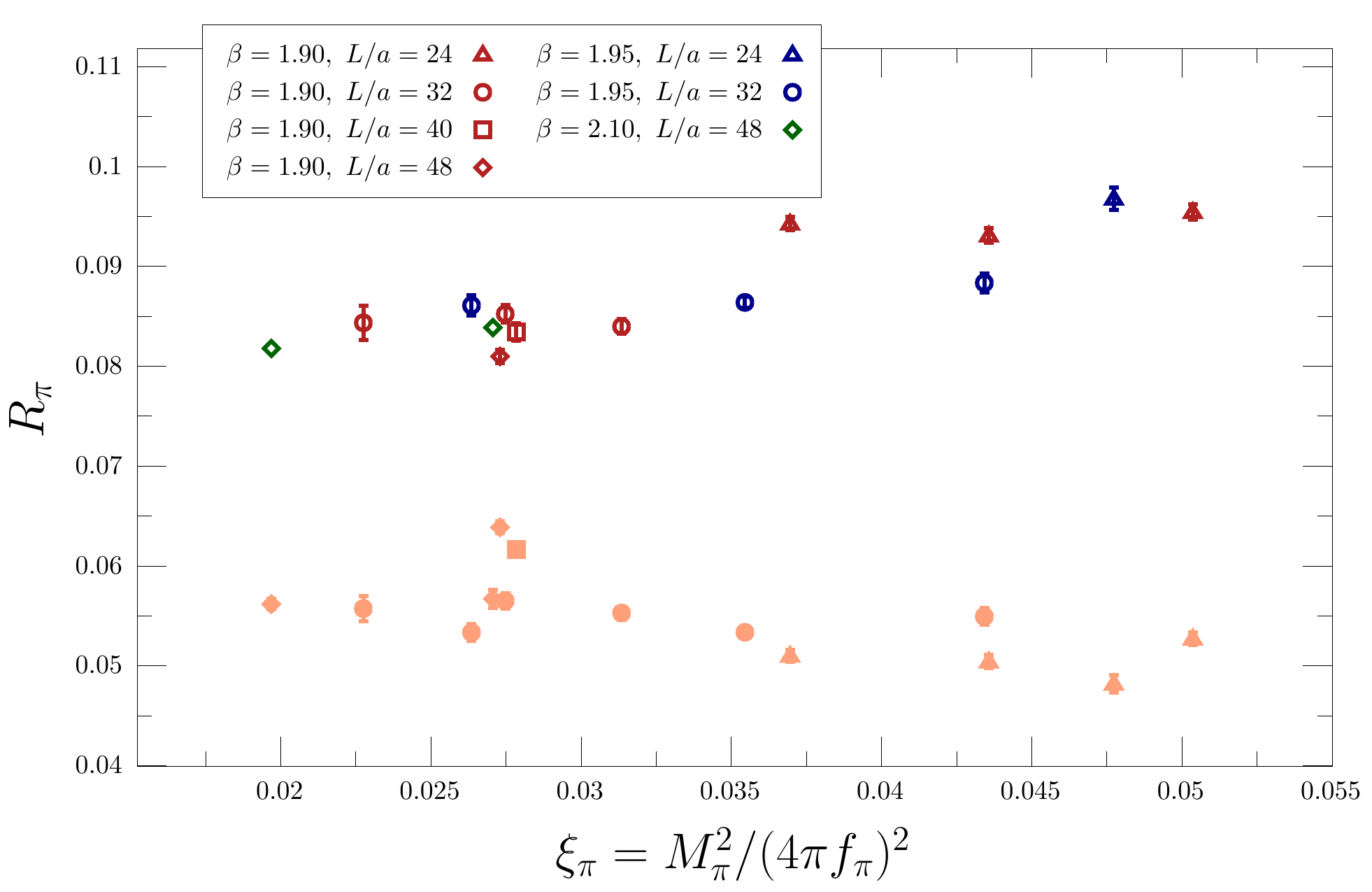}
    \caption{\small\it Our estimate of the ratio $R_{\pi}$ as a function of $\xi_{\pi}$, as determined in the RTM scheme and including the contribution of the disconnected diagram. The filled markers represent the data without any FSE correction, while the empty ones represent the result of the subtraction of the universal FSEs using Eq.~(\ref{eq:universal_FSE}).}
    \label{fig:QED2}
\end{figure}

The presence of the QED interactions, generate sizable FSEs in $M_{\pi^{+}}-M_{\pi^{0}}$ which are power-law in the spatial lattice extent $L$. The leading and next-to-leading FSEs of order $\mathcal{O}(1/L)$ and $\mathcal{O}( (1/L)^{2})$ are however universal~\cite{Borsanyi:2014jba, Davoudi:2014qua}, and in the case of the $QED_L$ used in this work, and for a pseudo-scalar meson of electric charge $Q$ and mass $M_{PS}$, are given by
\begin{equation}
M_{PS}^2(L) - M_{PS}^2(\infty) = - Q^2 \alpha_{em} \frac{\kappa}{L^2}\left( 2 +  M_{PS} L \right) ~ ,
     \label{eq:universal_FSE}
\end{equation}
where $\kappa=2.837297$. These corrections can be subtracted exactly from our lattice data leaving residual structure-dependent (SD) $\mathcal{O}((1/L)^{3})$ FSEs, as shown in Fig.~\ref{fig:QED1} for the ensembles of type $A40.XX$, which only differ in the spatial extent. In Fig.~\ref{fig:QED2} we show instead our determination of the $R_{\pi}$ ratio, before and after removal of the universal FSEs, and for all ensembles considered in this work, as a function of the dimensionless parameter $\xi_{\pi} = \frac{M_{\pi}^{2}}{(4\pi f_{\pi})^{2}}$. \\

Inspired by the ChPT analysis of Ref.~\cite{Hayakawa:2008an}, and by the non-relativistic expansion of Ref.~\cite{Davoudi:2014qua}, we extrapolated towards the physical pion mass and towards the continuum and infinite volume limit, employing the following Ansatz for the ratio $R_{\pi}$

\begin{align}
R_{\pi}^{\textrm{sub.}}(\xi_{\pi}, a, L) &=  4e^{2}C -3e^{2} \xi_{\pi}\log{\xi_{\pi}} + e^{2}A_{1} \xi_{\pi}+ e^{2}A_2 \xi_{\pi}^{2} +  e^{2}D a^2  \nonumber \\[6pt]
	 				    &+  e^{2}D_m \xi_{\pi} a^2 + e^{2}K \frac{(4\pi)^{2}\xi_{\pi}}{3M_{\pi}L^3} \langle r^2 \rangle_{\pi^+}  + e^{2}F_{a}\frac{\xi_{\pi}}{M_{\pi}}\frac{a^{2}}{L^{3}} ~,
    \label{eq:pion_fit}
\end{align}
where $R_{\pi}^{\textrm{sub.}}$ is the $R_{\pi}$ ratio after the subtraction of the universal FSEs, and $\langle r^2\rangle_{\pi^+} = (0.672 \pm 0.008 ~ \mbox{fm})^2$ is the squared pion charge radius. In the previous expression $C, A_{1}, A_{2}, D, D_{m}, K$ and $F_{a}$ are free fitting parameters. In particular $C$ and $A_{1}$ parameterize the ChPT expansion for $R_{\pi}$ up to NLO, $A_{2}$ is an effective LEC at NNLO, while $D$ and $D_{m}$ take into account discretization effects. Finally, $K$ takes into account deviations from the non-relativistic prediction of Ref.~\cite{Davoudi:2014qua} ($K=1$) for the SD FSEs, while the term proportional to $a^{2}/L^{3}$ corresponds to an expected FSE due to an heavy intermediate state of mass $\propto 1/a$~\cite{Tantalo:2016vxk}. \\
\begin{figure}
    \centering
    \includegraphics[scale=0.50]{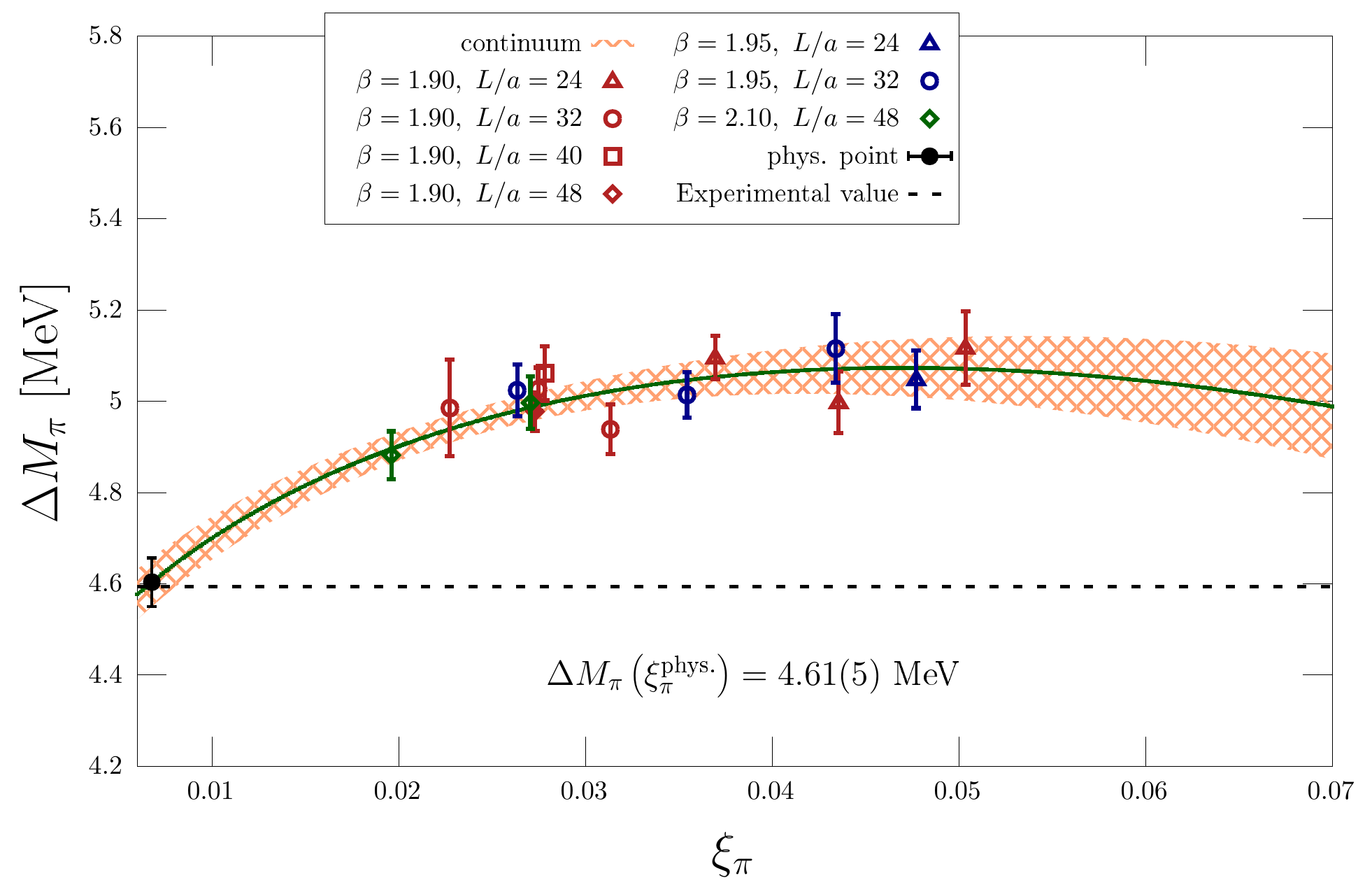}
    \caption{\small\it Result of the combined continuum/thermodynamic and physical point extrapolation as obtained setting $A_{2} = D = D_m = F_{a}= 0$. The data points correspond to our lattice data after the subtraction of the universal and of the SD FSEs, while the orange band sbows the statistical uncertainty after the continuum and infinite volume extrapolation. The black point corresponds to our determination at the physical point.}
    \label{fig:fit_QED}
\end{figure}

In Fig.~\ref{fig:fit_QED}, we show the result of the extrapolation obtained using the Ansatz of Eq.~(\ref{eq:pion_fit}), setting $A_{2} = D = D_m = F_{a}= 0$, which represents our preferred fit. The quantity $\Delta M_{\pi}$, which at the physical point gives the pion mass splitting, is defined as $
\Delta M_{\pi} \equiv R_{\pi}\cdot(f_{\pi}^{\rm{phys.}})^{2}/(2M_{\pi}^{\rm{phys.}})$,
with $M_{\pi}^{\rm{phys.}} = 134.977~{\rm MeV}$ and $f_{\pi}^{\rm{phys.}} = 130.4~{\rm MeV}$.  
Notice the remarkable smallness of $\mathcal{O}(a^{2})$ lattice artifacts in our data, within accuracy. To estimate systematic errors, we performed a total of $24$ fits, differing on whether the $A_{2}$, the $D$, and the $D_{m}$ fit parameters have been included or not, and on the form of the SD FVEs for which we either include $K$ or $F_{a}$ as a free fit parameter (in this last case setting $K=1$), or introduce an additional $1/L^{4}$ term on top of the non-relativistic prediction $K=1, F_{a}=0$. The fit results have been combined using the Akaike Information Criterium (AIC), in which, to each fit, it is assigned a weight $w_{i} \propto \exp{-(\chi^{2}+2n_{pars})/2}$.
Mean values and standard errors have been then computed making use of Eqs.~(38)-(43) from Ref.~\cite{alexandrou2021ratio}. 
Our preliminary result for the pion mass splitting is
\begin{align}
M_{\pi^{+}}- M_{\pi^{0}} = 4.622~(64)_{stat.}(70)_{syst.}~{\rm MeV} = 4.622~(95)~{\rm MeV}~,
\end{align}
which agrees very well with the experimental determination $
\left[ M_{\pi^{+}}- M_{\pi^{0}}\right]^{\emph{exp.}} \,=\, 4.5936~(5)~ \textrm{MeV}$, 
and with the result of a recent lattice determination~\cite{Feng:2021zek} $M_{\pi^+} - M_{\pi^0} = 4.534(42)(43)~{\rm MeV}$,  in which the disconnected contribution has been computed as well.  
\section{Conclusions}
We have presented an analysis of the $\mathcal{O}(\alpha_{em})$ and $\mathcal{O}( (m_{d} -m_{u})^{2})$ mass splitting $M_{\pi^{+}}-M_{\pi^{0}}$ between the charged and neutral pion. We showed that a good accuracy in the determination of the disconnected diagrams can be achieved by working in the \textit{rotated twisted mass} (RTM) scheme, which have been shown to be particularly convenient for the evaluation of some QCD$+$QED mesonic observables based on the RM123 approach.
By evaluating the strong IB contribution to $M_{\pi^{+}}- M_{\pi^{0}}$ at $\mathcal{O}( (m_{d}-m_{u})^{2})$ and the coupling $Z_{P^{0}\pi^{0}}$ of the neutral pion to the isoscalar operator at order $\mathcal{O}(m_{d}-m_{u})$, we showed that it is possible to cleanly extract the value of the ChPT $\rm{SU}(2)$ LEC $\ell_{7}$. Our results for $\ell_{7}$~(\ref{eq:res_l7}) is in agreement with previous estimates while improving significantly the precision.
For the $\mathcal{O}(\alpha_{em})$ mass splitting $M_{\pi^{+}}-M_{\pi^{0}}$, after extrapolating to the continuum and infinite volume limit, and at the physical point, our preliminary estimate~(\ref{eq:prelim_res_QED}) has a total uncertainty of $\sim 2\%$ and agrees with the experimental result. 
\section{Acknowledgement}
We thank C. Tarantino for useful discussions, and all members of the ETMC for the most enjoyable collaboration. We acknowledge CINECA for the provision
of CPU time under the specific initiative INFN-LQCD123 and IscrB\_S-EPIC. F.S. G.G and S.S. are supported by
the Italian Ministry of University and Research (MIUR) under grant PRIN20172LNEEZ. F.S. and G.G are supported by INFN under GRANT73/CALAT. R.F. acknowledges support from the University of Tor Vergata through the Grant “Strong Interactions: from Lattice QCD to Strings, Branes and Holography” within the Excellence Scheme “Beyond the Borders”.

\end{document}